\documentclass[twocolumn]{IEEEtran}
\ifCLASSOPTIONcompsoc
  \usepackage[nocompress]{cite}
\else
  \usepackage{cite}
\fi

\ifCLASSINFOpdf
   \usepackage[pdftex]{graphicx}
\else
\fi
\usepackage{xcolor}
\usepackage{physics}
\usepackage{ulem}
\usepackage{placeins}
\renewcommand{\baselinestretch}{1}

\begin{document}

\title{A Set of Benchmark Tests for Validation of 3D Particle In Cell Methods}
\author{S. O'Connor,~\IEEEmembership{Student Member,~IEEE,}
        Z. D. Crawford,~\IEEEmembership{Student Member,~IEEE,}
        J. Verboncoeur,~\IEEEmembership{Fellow,~IEEE,}
        J. Lugisland,~\IEEEmembership{Fellow,~IEEE,}
        B. Shanker,~\IEEEmembership{Fellow,~IEEE}
\IEEEcompsocitemizethanks{\IEEEcompsocthanksitem  S. O'Connor, Z. D. Crawford, J. Verboncour, J. Lugisland, B. Shanker are with the Department
of Electrical and Computer Engineering, Michigan State University, East Lansing,
MI, 48824.\protect\\
E-mail: oconn220@msu.edu}
\thanks{}}

\markboth{}%
{Shell \MakeLowercase{\textit{et al.}}: Bare Demo of IEEEtran.cls for Computer Society Journals}
\IEEEtitleabstractindextext{%
\begin{abstract}
While the particle-in-cell (PIC) method is quite mature, verification and validation of both newly developed methods and individual codes has largely focused on an idiosyncratic choice of a few test cases. 
Many of these test cases involve either one- or two-dimensional simulations. 
This is either due to availability of (quasi) analytic solutions or historical reasons.
Additionally, tests often focus on investigation of particular physics problems, such as particle emission or collisions, and do not necessarily study the combined impact of the suite of algorithms necessary for a full featured PIC code.
As three dimensional (3D) codes become the norm, there is a lack of benchmarks test that can establish the validity of these codes; existing papers either do not delve into the details of the numerical experiment or provide other measurable numeric metrics (such as noise) that are outcomes of the simulation. 

This paper seeks to provide several test cases that can be used for validation and bench-marking of particle in cell codes in 3D. 
We focus on examples that are collisionless, and can be run with a reasonable amount of computational power. 

Four test cases are presented in significant detail; these include, basic particle motion, beam expansion, adiabatic expansion of plasma, and two stream instability. All presented cases are compared either against existing analytical data or other codes. 

We anticipate that these cases should help fill the void of bench-marking and validation problems and help the development of new particle in cell codes. 
\end{abstract}

\begin{IEEEkeywords}
Particle in Cell, PIC, Finite Element Method, Maxwell Solvers, Vlasov Equations, Plasma, Space Charge
\end{IEEEkeywords}}

\maketitle

\IEEEdisplaynontitleabstractindextext

\IEEEpeerreviewmaketitle

\section{Introduction}\label{sec:introduction}

 \IEEEPARstart{S}{imulation} of space charge and plasma is an integral part of many scientific and engineering processes, with specific applications of pulsed power, particle accelerators, directed energy, integrated chip manufacture, satellites, and medicine, to name a few examples. 
A number of methods have been used to simulate the underlying physics; these range from molecular dynamics, many-body, global-model, 2-fluid, magneto-hydrodynamics (MHD), etc. \cite{chen2012introduction}. 
The key with all of these methods is the self-consistent evolution of the fields and plasma. 
In the area of kinetic computational plasma physics, this means evolving the particle distribution function in a manner consistent with dynamic electric and magnetic fields.
One such method, the Particle-in-Cell (PIC), does this by (a) mapping all particle-field interactions through a mesh, and (b) using macro-particles as statistically significant markers of the distribution function's phasespace   \cite{birdsall2018plasma,hockney1988computer}.
PIC has been around for several decades, originating in the 1960's.
Much of the early work focused on using finite-difference time-domain (FDTD) for electromagnetic fields solutions on a regular grid, more specifically, Yee Cells \cite{yee1966numerical}. 
This method has become popular due to its simplicity and scalabilty \cite{qiang2000object}.  
Additionally, these methods are local, which makes parallel implementation relatively straight forward. 
While efficient, curved surfaces, or any geometry that is not aligned with a regular structured grid, is poorly represented.
Significant work has been done to adapt Cartesian grids with better model boundaries. 
The most popular is the Dey-Mittra adaption of FDTD to PIC \cite{dey1997technique}. 
It should be noted that there have been a number of other advances in conformal FDTD\cite{meierbachtol2015conformal}, but they have not made their way into PIC codes. 
An alternative is to use  an  axisymmetric methods when possible by treating a three dimensional problem in a two-dimensional setting.
An excellent example is XOOPIC, which is  open source and has been used extensively \cite{verboncoeur1995object}.

Since the introduction of FDTD, the fields community has invested time and resources to developing high fidelity field solvers. 
These include finite element method (FEM) and discontinuous Galerkin methods (DG). 
Both can provide higher order representation of fields and geometry. 
Additionally, one has better understanding of efficient time marching schemes \cite{jin2015finite}. 
Here, our focus is finite element based Maxwell solver, that have been a) adapted to PIC and b) shown to be devoid of several problems associated with charge conservation.

The challenge with using unstructured grids, using FEM or DG, is developing a consistent scheme to solve the coupled problem of evolving fields and particle population. 
Given the relative lack of maturity of FEM-PIC, it is important to understand how the nuances of these algorithms affect the accuracy of the physics that is predicted. 
Small variations in method can have significantly different prediction; for instance, time stepping method, solving the first order Maxwell's equations vs. solving the second order wave equation, order of spatial basis functions, order of time basis function, etc., can yield different predictions.
Perhaps, the most critical of these nuances is charge conservation in that it explicitly ties together the evolution of the fields and the particles.  
Additionally, while there has been extensive efforts for FEM in the context of source-free electromagnetics, FEM-PIC is much less developed. 
The most recent of these is the work in \cite{pinto2014charge} where a charge conserving scheme for Whitney basis was presented. 
This has been developed further by Teixeira  \cite{moon2015exact}.

This begs the question as to how PIC codes are validated. 
Correctness of a simulation has typically been done through a number of methods such as simple property test such as energy conservation, to more rigorous tests such as rates of convergence. These test are described in more rigor in \cite{riva2017methodology}. 
The most rigorous tests require a comparison with analytical solutions. 
But, this is much harder to come by in PIC settings, as analytic solution for most problems do not exist. 
Most tests rely either demonstrating proper behavior, or comparison to another code. 
As a result, there is a need for a set of test examples that can be used to validate/benchmark and see how these variations affect the solution of the solver. 
Validation and verification methods play import roles in quantifying error, numerical noise, and comparison of method performance. 
Of late, in the PIC community, there has been a push for more robust bench marking and validation of (PIC) codes
\cite{turner2013simulation,turner2016verification,riva2017methodology}. 
Much of this current literature focuses on discharge and particle emission and collisions. 
Less attention is paid to test examples where the grid-particle interaction dominate the numerical effects. 
This is the void we seek to fill. 

In collision-less PIC, there are a number of metrics one may want to measure such as field noise, charge conservation and energy fluctuation. A set of benchmark tests are needed to measure each of these metrics. 
In this paper, we have devised a set of test cases to measure various metrics of particle in cell codes to both verify the code and allow for comparisons between codes. 
The four cases presented here have (quasi)-analytic solutions, and can be verified against other methods such as an electrostatic PIC codes.
We provide details of each case such that comparison and validation is easy. 
Each of these cases are a 3-D example that can be set up with simple boundary conditions -- perfect electric conductor (PEC) or first order absorbing boundary condition (ABC). 
The domain size to capture the physic of the problem is limited such that large amount of computational resources are not needed, but can be scaled in a parallel context, to test algorithm scaling.

\section{Problem Description}
Next, we present a succinct outline of the problem description and implementation details for the test cases presented.
Consider a domain $\Omega$ enclosed by a boundary $\partial \Omega$. 
The domain consists of a free space permittivity and permeability defined by $\epsilon_0$ and $\mu_0$, and can contain multiple charge species. 
Both the fields and charge distribution evolve over time self consistently with accordance to both the Vlasov Eq. (\ref{eq:vlasov}) and Maxwell's Eq. (\ref{eq:maxwell}),
\begin{align} \label{eq:vlasov}
  \partial_t f(t,\vb{r},\vb{v})  + \vb{v} \cdot \nabla f(t,\vb{r},\vb{v}) + \\ \frac{q}{m} [\vb{E}(t,\vb{r}) + \vb{v} \times \vb{B}(t,\vb{r})] \cdot \nabla_v f(t,\vb{r},\vb{v}) = 0. \nonumber
\end{align}
Here, $f(t,\vb{r},\vb{v})$ is the phase space distribution function. Typically, one does not solve Eq. $(\ref{eq:vlasov})$ directly. 
The usual approach, that we follow as well, is to define charge and current via the first and second moment of this distribution function;
i.e., $\rho(t,\vb{r})=q\int_{\Omega} f(t,\vb{r},\vb{v})d\vb{v}$ and $\vb{J}(t,\vb{r})=q\int _{\Omega}\vb{v}(t)f(t,\vb{r},\vb{v})d\vb{v}$.
These are then solved self-consistently with Eq. (\ref{eq:maxwell}) and the particle location are updated using the Lorentz force $\vb{F} = q(\vb{E} + \vb{v}\times \vb{B})$.

\begin{subequations}\label{eq:maxwell}
    \begin{equation}
        - \frac{\partial \vb{B}(t,\vb{r})}{\partial t} = \curl \vb{E}(t,\vb{r})
    \end{equation}
    \begin{equation}
        \frac{\partial \vb{D}(t,\vb{r})}{\partial t} = \curl \vb{H}(t,\vb{r}) - \vb{J}(t,\vb{r})
    \end{equation}
    \begin{equation}
        \div \vb{B}(t,\vb{r}) = 0
    \end{equation}
    \begin{equation}
        \div \vb{D}(t,\vb{r}) = \rho (t,\vb{r})
    \end{equation}
\end{subequations}
We assume that the boundary can be subdivided into either Dirichlet $\Gamma_D$, Nuemann $\Gamma_N$ or impedance $\Gamma_I$, boundary conditions as defined by,
\begin{subequations}\label{eq:bceq}
\begin{equation}
\hat{n}\times \mathbf{E}(\mathbf{r},t) = \mathbf{\Psi}_D(\mathbf{r},t)\;\;\text{on}\;\Gamma_D,
\end{equation}
\begin{equation}
\hat{n}\times \frac{\mathbf{B}(\mathbf{r},t)}{\mu} = \mathbf{\Psi}_N(\mathbf{r},t)\;\;\text{on}\;\Gamma_N,
\end{equation}
\begin{equation}
\hat{n}\times \frac{\mathbf{B}(\mathbf{r},t)}{\mu} - Y\hat{n}\times\hat{n}\times \mathbf{E}(\mathbf{r},t) = \mathbf{\Psi}_I(\mathbf{r},t)\;\;\text{on}\;\Gamma_I.
\end{equation}
\end{subequations}
where $\hat{n}$ is an outward pointing normal to $\Gamma_D$, $\Gamma_N$ or $\Gamma_I$.
We define $Y = \sqrt{\epsilon_0/\mu_0}$ as the free space surface admittance, and $ \mathbf{\Psi}_D(\mathbf{r},t), \mathbf{\Psi}_N(\mathbf{r},t),\mathbf{\Psi}_I(\mathbf{r},t)$ are boundary condition functions.
Initial conditions, must satisfy Gauss's electric and magnetic law, such that $\div \vb{D} (0,\vb{r}) = \rho (0, \vb{r})$ and $\div \vb{B} (0,\vb{r}) = 0$. 
The algorithm's preservation of these conditions can then be tested by our problems.

\subsection{The Discrete Domain}
To solve the system, we use a 3D finite element Maxwell solver.
Consider a domain $\Omega$ broken into an irregular non-overlapping tetrahedral elements with boundaries $\partial \Omega$.
In each tetrahedral element the field quantities $\vb{E}$ and $\vb{B}$ can be represented using Whitney spaces.
The moments of the distribution function can be represented using using delta functions, shown in  Eq. (\ref{eq:dcont_charge}),(\ref{eq:dcont_current}),
\begin{subequations}
    \begin{equation}\label{eq:dcont_charge}
        \rho_\alpha(t,\vb{r}) = q_{\alpha}\sum_p^{N_p} \delta(\vb{r}-\vb{r}_p(t))
    \end{equation}            
    \begin{equation}\label{eq:dcont_current}
        \vb{J}_\alpha(t,\vb{r}) = q_{\alpha}\sum_p^{N_p}\vb{v}_p(t)\delta(\vb{r}-\vb{r}_p(t))
    \end{equation}
\end{subequations}
for each charge species $\alpha$, where $q_{\alpha}$ is the charge of the macro-particle and $N_p$ is the number of macro-particles.
The system proceeds as follows.
The Whitney basis functions are used to convert Maxwell's continuous equation to discrete ones.
Current and charge are mapped to the grid via Whitney-1 $\big(\vb{W}^{(1)}_i\big)$ and Whitney-0 $\big(W^{(0)}_j\big)$ basis sets, and the solution to Faraday's and Ampere's law uses a leap frog scheme. 
Details to pertinent to each stage of the scheme can be found in \cite{pinto2014charge,moon2015exact}.
Using $ \vb{E}(t,\vb{r}) = \sum_{i=1}^{N_e} e_i(t) \vb{W}^{(1)}_{i}(\vb{r}) $ and  $ \vb{B}(t,\vb{r}) = \sum_{i=1}^{N_f} b_i(t) \vb{W}^{(2)}_{i}(\vb{r}) $ and Galerkin testing results in, 
\begin{align}
\vb{B}^{n+1/2} &=\vb{B}^{n-1/2} - \Delta_t [D_{curl}]\cdot\vb{E}^n \label{eq:faraday}\\
\vb{E}^{n+1} &= \vb{E}^{n} + \Delta_t[\star_{\epsilon}]^{-1}\cdot\bigg( [D_{curl}]\cdot[\star_{\mu^{-1}}] \cdot \vb{B}^{n+1/2} - \vb{J}^{n+1/2}\bigg) \label{eq:ampere}
\end{align}
where $N_e$ and $N_f$ are the number of edges and faces, $\Delta_t$ is the time step size,
\begin{align*}
    \vb{E}^{n} &= [e^{n}_1,e^{n}_2,...,e^{n}_{N_e}], \quad \\
    \vb{B}^{n+1/2} &= [b^{n+1/2}_1,e^{n+1/2}_2,...,b^{n+1/2}_{N_f}], \quad \\
    \vb{J}^{n+1/2} &= [j_1^{n+1/2},j_2^{n+1/2},...,j_{N_e}^{n+1/2}]
\end{align*}
with $\vb{E}^n$, $\vb{B}^{n+1/2}$, and $\vb{J}^{n+1/2}$ being the field coefficients at time step $n$ and $n+1/2$. 
The Hodge and discrete curl matrices in  Eq. (\ref{eq:faraday}) and Eq. (\ref{eq:ampere}) are,
\begin{equation}
        [\star_\epsilon]_{i,j} = \langle \vb{W}^{(1)}_i(\vb{r}),\varepsilon\cdot\vb{W}^{(1)}_j(\vb{r}) \rangle
\end{equation}
\begin{equation}
        [\star_{\mu^{-1}}]_{i,j} = \langle \vb{W}^{(2)}_i(\vb{r}),\mu^{-1}\cdot\vb{W}^{(2)}_j(\vb{r})\rangle
\end{equation}
\begin{equation}
        [D_{curl}]_{i,j} = \langle \vb{\hat{n}}_i , \curl \vb{W}^{(1)}_j(\vb{r})\rangle
\end{equation}
where, $[D_{curl}]$ is the discrete curl matrix, $[\star_{\mu^{-1}}]$ is the Hodge star matrix, $[\star_\epsilon]$ is the Hodge epsilon matrix.
The current is mapped using a charge conserving method presented in \cite{pinto2014charge,moon2015exact},
\begin{align}
    j_i^{n+1/2} &= \frac{q_p}{\Delta_t}  \int_{\vb{r}^{n}_p}^{\vb{r}^{n+1}_{p}} \vb{W}^{(1)}_{i}(\vb{r}) \cdot d\vb{l} \\
    &=\frac{q_p}{\Delta_t}[W^{(0)}_i(\vb{r}^{n}_p)W^{(0)}_j(\vb{r}^{n+1}_p) - W^{(0)}_i(\vb{r}^{n+1}_p) W^{(0)}_j(\vb{r}^{n}_p)],
\end{align}
where $W^{(0)}_i(\vb{r})$ is the barycentric coordinate for node $i$ within a cell.
\section{Benchmark Tests}
In this section, we present four cases that could be used for validation. 
In each of these cases, we will present a number of metrics and discuss what would happen if some of these metrics were not to be satisfied.
Each test is designed to highlight an aspect of the method. 
The first test case, basic particle motion, highlights the particle push's performance.
The expanding particle beam is setup to verify current mapping, discrete Gauss's magnetic and electric law, and the discrete continuity equation.
The adiabatic expanding plasma test case allows for multi-species evolving distributions and can be scaled up to test the scalability of methods.
Two stream instability displays more dynamic features of the method and has analytic solution based on the rise time of the simulation.

\subsection{Basic Particle Motion}
The first tests of any PIC method should start with validating the particle motion. 
To this end, we set up two tests, 1) a particle moves under the influence of an external field due to Newton's laws and 2) the motion of a particle is tracked under the influence of both a background external field and grid interaction solved by Maxwell's equation and Newton's laws. 
We use Newton's law to obtain position and velocity at every time point. 
Such a test verified that the trajectory of the particle --- as mapped on a grid --- is computed correctly. 
Given that this trajectory can be obtained analytically, one can obtain error bounds and relate them to time step size. 

\begin{figure}
    \centering
    \includegraphics[draft=false,scale=0.55]{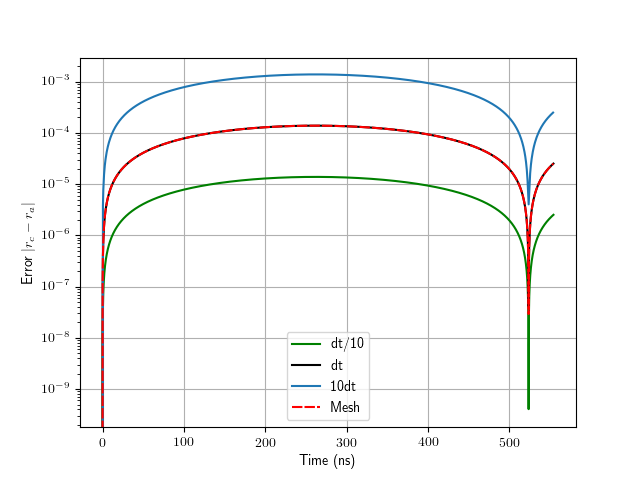}
    \caption{Cyclotron motion error from just Boris push with several different time step sizes, as well as influence from the mesh.}
    \label{fig:BasicMotionBFieldError}
\end{figure}

Next this test can be rerun assuming radiation, that is solving Newton and Maxwell's equation self consistently. 
If the solutions are consistent, the two path should be approximately identical as the radiated fields are significantly smaller than the DC fields. 
All results that bear this exertion are shown in Fig \ref{fig:BasicMotionBFieldError}. 

The parameters chosen are such that force due to the particle motion is significantly less than the force due to the DC background fields. 
This choice causes the main source of error in the particle trajectories to be dominated by the particle push and not the field solver. 
This setup also presents an opportunity to verify current continuity and the discrete Gauss's law in a setup where the particle path is for all intents and purposes is known.
Although the field values are small, they are non-zero. 
This case is simple enough that it can be performed on a small enough meshes that any errors are easily identified.

\begin{figure}
    \centering
    \includegraphics[draft=false,scale=0.55]{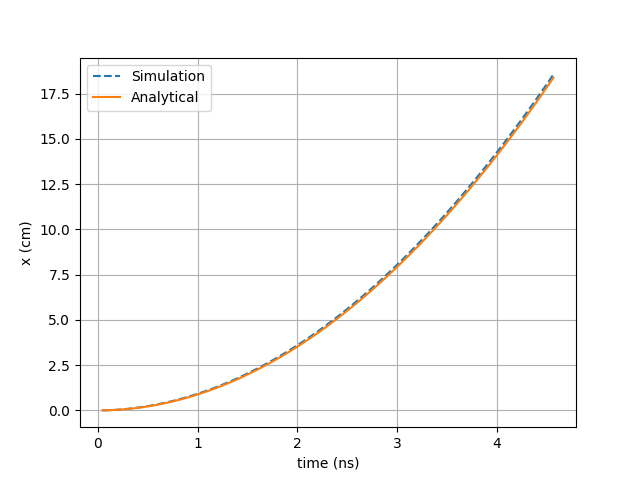}
    \caption{Particle acceleration due to electric field.}
    \label{fig:BasicMotionEField}
\end{figure}

\begin{figure}
    \centering
    \includegraphics[draft=false,scale=0.55]{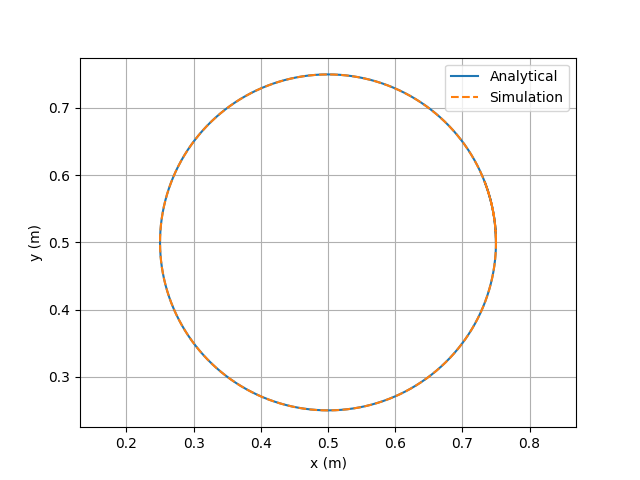}
    \caption{Cyclotron particle motion of an electron in a magnetic field.}
    \label{fig:BasicMotionBField}
\end{figure}

\begin{figure}
    \centering
    \includegraphics[draft=false,scale=0.55]{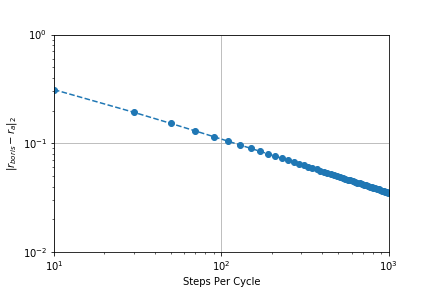}
    \caption{Error as time step size decreases for Boris push.}
    \label{fig:boris_error}
\end{figure}

\paragraph{External E-Field}    
The first term of the Lorentz force can be checked by placing a particle at rest in an uniform background electric field of $1$ V/m and letting the particle accelerate as shown in Fig. \ref{fig:BasicMotionEField}.
The macro-particle has the mass and charge of a single electron.
We calculate the analytic motion and the error we get from the particle push method.
\paragraph{External B-Field}    
To check the $\vb{v} \cross \vb{B}$ term we, apply a uniform external magnetic flux density $\vb{B}=6.82272\cdot10^{-5}\hat{z}$ $Wb/m^2$ with no electric field. 
A single particle with the mass and charge of an single electron is given an initial velocity $\vb{v}=3\cdot 10^6 \hat{y}$ m/s and position $\vb{r}=[0.75,0.5,0]$ which results in a gyro-radius of $0.5$m, as shown in Fig. \ref{fig:BasicMotionBField}.
We analyze the error in Fig. \ref{fig:BasicMotionBFieldError} by comparing four test cases; the motion due to a fully coupled system labeled $mesh$, the background fields with the current time step size, background fields with the time step size 10 time large, and background fields with the time step 10 time smaller. 
The error is calculated by finding the distance between the analytic and simulated locations at a given step for each case.
The two dips in the error stem from the particle starting at the same location --- zero error --- and drifting array from the analytic circle and once the rotation is complete ending up at the start location. 
In Fig. \ref{fig:boris_error}, we compare the average error over one cycle of cyclotron motion vs. the number of step per cycle. 

\FloatBarrier

\subsection{Expanding Particle Beam}
In the second test, a beam of electrons are injected into a cylindrical drift tube and is evolved over time. 
The reasons for this test are 1) it has an approximate solution found in \cite{reiser1994theory};
2) charge conservation is easy to measure and any error shows up in the solution as higher noise levels or spatial striations in the beam \cite{barker2001high}. 

\begin{figure}
    \centering
    \includegraphics[draft=false,scale=0.55]{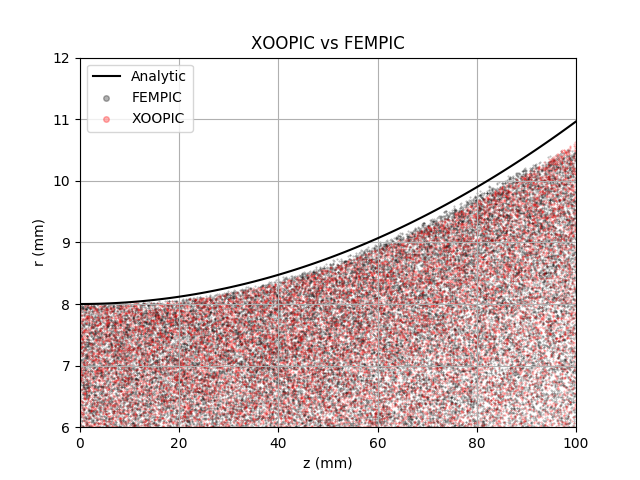}
    \caption{XOOPIC vs FEMPIC for use in particle in cell }
    \label{fig:xoopic_vs_fem_PIC}
\end{figure}

We create a cylindrical drift tube geometry as shown in Fig. \ref{fig:xoopic_vs_fem_PIC}, with a radius of $0.02$ m and a length of $0.1$ m. We set all exterior surfaces to perfect conductors (PEC) with the interior volume set to vacuum. 
The drift tube is discretized such that the average edge length is $4.11$ mm. 
The beam has a voltage and current of 7.107 kV and 0.25 A. 
This setup amounts to a beam radius of $8$ mm centered in the cylinder with a starting velocity of $v_0=[0.0,0.0,5 \cdot 10^7]$ m/s. 
The beam is ramped up by changing the macroparticle weight from 0 to 52012 electrons per macroparticle linearly over the first 2 ns of the simulation. 
The parameters are listed in Table \ref{tb:beam}.
\begin{table}[ht!]
\centering
    \caption{Expanding Particle Beam Parameters}
    \begin{tabular}{c|c}
         \textit{Parameter} & \textit{Value}  \\
         \hline
         Cavity Radius & 20 mm \\
         Cavity Length & 100 mm\\
         Boundary Conditions & PEC \\
         $v_p$ & $5\cdot10^7$ m/s \\ 
         $v_p/c$ & 0.16678 \\ 
         beam radius $r_b$ & 8.00 mm \\
         Number particles per time step & 10 \\
         species & electrons \\
         Turn on time & 2 ns \\
         beam current  & 0.25 A \\
         macro-particle size & 52012.58 \\
         min edge length & 1.529 mm \\
         max edge length & 6.872 mm \\
    \end{tabular} \label{tb:beam}
\end{table}
The current in this example was chosen such that a converged solutions of the expansion could be represented with a coarse mesh to reduced the simulation time needed to solve the problem. 
We also compare the results with the 2-D FDTD code XOOPIC in $r$-$\theta$ mode. As is evident, the agreement between the two codes --- despite obvious advantages afforded by asymmetry in XOOPIC --- agree well with each other.
The trajectories of the particle is shown in Fig. \ref{fig:xoopic_vs_fem_PIC}, and there is excellent agreement between FEMPIC and XOOPIC. 
While FEMPIC and XOOPIC agree with each other, they differ slightly from the analytic data, although there is good agreement on the overall trend. 
The analytic solution derived in \cite{reiser1994theory} is also shown in Fig. \ref{fig:xoopic_vs_fem_PIC},
it assumes an idealized beam, with uniform density radially and ignores the perfect electric conductor (PEC) boundary conditions on the entrance and exit surfaces of the beam. 
We postulate that this leads to the slight disagreement. 
Note, it is difficult, if not impossible to create the exactly identical conditions for either the numerical simulation or analytical analysis. 
Nerveless, we include this data to show that the same trends are captured. 

\begin{figure}
    \centering
    \includegraphics[draft=false,scale=0.5]{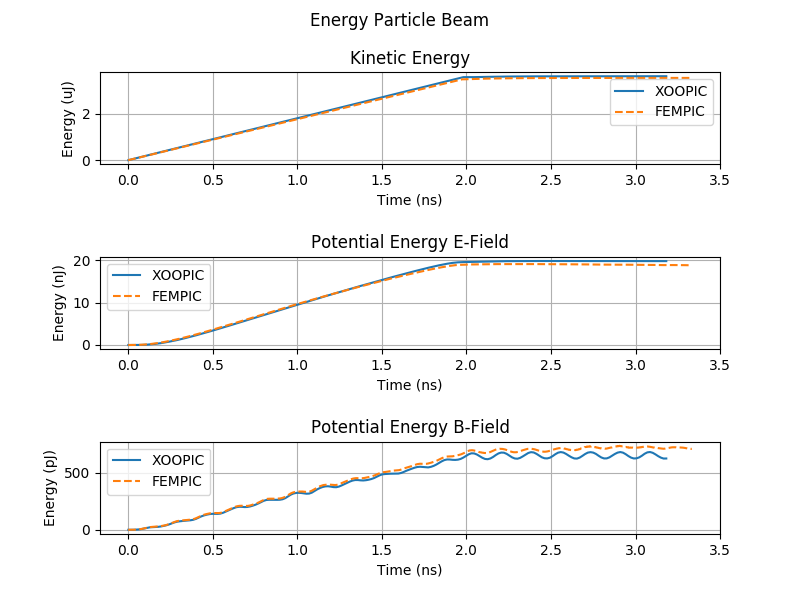}
    \caption{Particle Beam Energy with a 2ns turn on time}
    \label{fig:beam_energy}
\end{figure}

\begin{figure}
    \centering
    \includegraphics[draft=false,scale=0.55]{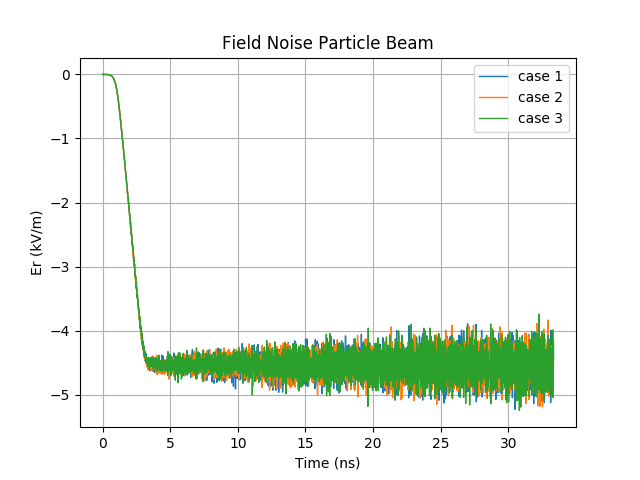}
    \caption{Noise measured from three beam test cases with different random number generator}
    \label{fig:beam_noise}
\end{figure}

\begin{figure}
    \centering
    \includegraphics[draft=false,scale=0.45]{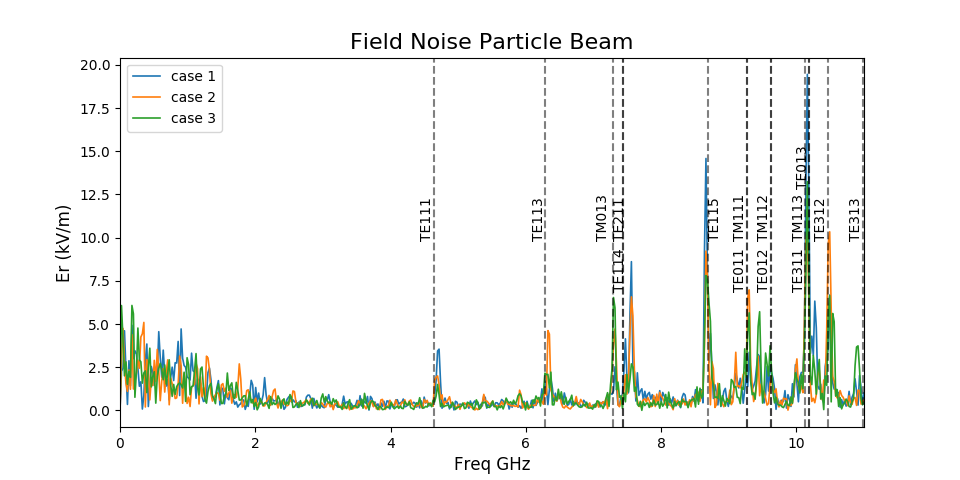}
    \caption{Fourier transform of the electric field for three beam tests that had different random number generator seeds.}
    \label{fig:beam_fft_noise}
\end{figure}

\begin{figure}
    \centering
    \includegraphics[draft=false,scale=0.55]{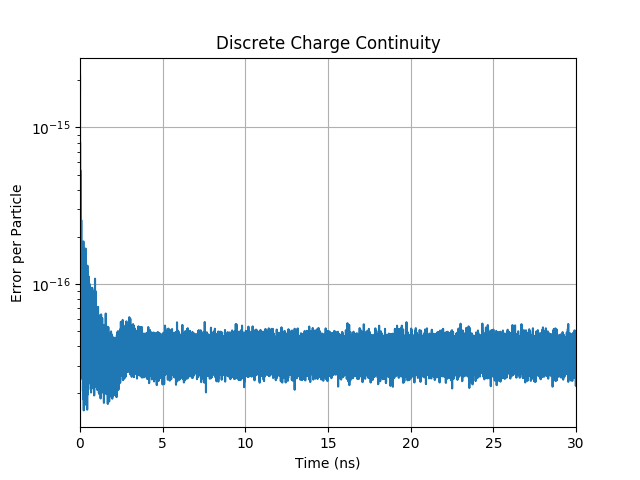}
    \caption{Charge Conservation for particle beam test case measured by $|(\rho^{n+1}-\rho^{n})/\Delta_t - [\nabla \cdot]J^{n+1/2}|/|(\rho^{n+1}-\rho^{n})/\Delta_t|/N_{part}$}
    \label{fig:charge_conservation}
\end{figure}

Next, the kinetic and potential energy  in both the electric and magnetic field obtained by FEMPIC and XOOPIC agree as shown in Fig. \ref{fig:beam_energy}.
We also examined field noise. 
Note, the boundaries are perfect electrically conducting, and there are no losses in this problem.
This results in cavity modes that are excited.
The radial component of the electric field, $E_r$, at halfway through the tube at a radial distance of $18$ cm from the center of the tube is shown for three cases in Fig. \ref{fig:beam_noise}.
Each case uses a different random number generator seed for the location of where the particle in the beam is injected. 
This allows us to see if the noise is random or does it fall into certain modes. 
Various modes of the cylindrical cavity are shown in Fig. \ref{fig:beam_fft_noise}. These show up upon taking the Fourier transform of the noise after the system reaches steady state at $6.66$ns and the DC component had been removed.
Non-random injection does artificially reduces the noise of the system.
We also measured the discrete continuity equation as a function of time in Fig. \ref{fig:charge_conservation}, it is evident charge is conserved to machine precision. 

\subsection{Adiabatic Expanding Plasmas}
In the third test, we simulate an adiabatically expanding plasma ball with a Gaussian distribution in the radial direction. This experiment has an approximate analytic solutions \cite{kovalev2003analytic} that can used for validation. 
These analytic solutions have been compared with experimental results in \cite{laha2007experimental}.
In earlier examples, the analytical examples are obtained in 1-D and 2-D, making it difficult to show a proper expansion with 3-D numerical code \cite{moon2015exact}. 
As analytical data is available in 3-D, 
we design the test to reflect the experiment done in \cite{laha2007experimental} but change the density such that the Debye length can fully resolved with modest computational resources. 
\begin{figure}
    \centering
    \includegraphics[draft=false,scale=0.55]{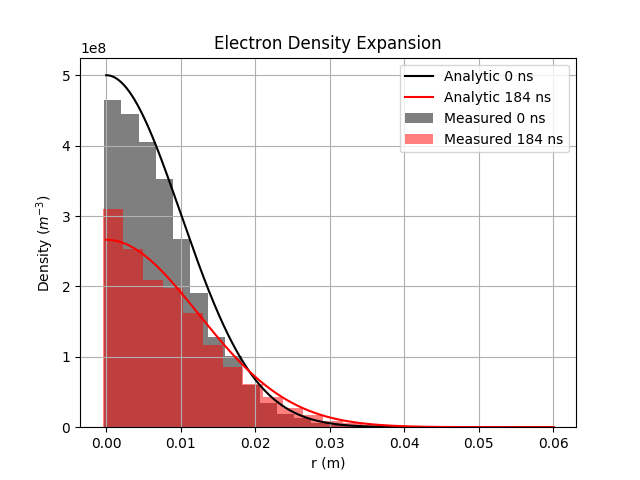}
    \caption{Distribution of electrons radially at two instants in time with analytic solutions.}
    \label{fig:plasma_ball_dis}
\end{figure}
The expanding plasma has two species, Sr+ ions with an initial temperature of 1 K and electrons at 100 K. 
The initial number density of both species is $5e8$ particle per $m^{-3}$, and the initial $\sigma=0.01$m such that the system is initial charge neutral everywhere. 
Other details of the simulation parameters can be found in Table \ref{tabe:plasma_ball}.
\begin{table}[]
    \centering
    \caption{Adiabatic Expanding Plasmas}
    \begin{tabular}{c|c} \label{tabe:plasma_ball}
         \textit{Parameter} & \textit{Value}  \\
         \hline
         Mesh Radius & 6mm \\
         Boundary Conditions & First order ABC \\
         $T_{ion}$ & 1K \\ 
         $T_{electron}$ & 100K \\ 
         Number Particles & 8000 \\
         Species & Electrons and $Sr^{+}$ \\
         Macro-Particle Size & 52012.58 \\
         Min Edge Length & 1.529mm \\
         Max Edge Length & 6.872mm \\
    \end{tabular}
\end{table}
When the simulation starts, the electrons expand outward faster than ions, due to the higher temperature, creating a radial electric field.
After the initial expansion outward, the growing electric field slows the expansion of the electrons. 
At the same time the ions expand outward pulled by the electrons.
This example can be scaled -- by changing the density of charges -- such that various levels of computational power are needed thus making it a good test of the scalability of a code.  
We use an first order absorbing boundary condition (ABC) to truncate the domain and place the boundary conditions sufficiently far from the initial distribution to capture the initial expansion of the electrons but not exit the mesh. 
Early electron expansion can be validated in small geometries, which we do here. 
The analytic distribution of electrons at initial conditions and at time 168 ns, are show in Fig. \ref{fig:plasma_ball_dis}.
This was derived in \cite{laha2007experimental}, and is succinctly presented here. 
The density function $f_{\alpha}$ of species $\alpha$ can be expanded as, 
\begin{equation}
 f_{\alpha}(t,\vb{r},\vb{v})=n_0 \bigg[\frac{m_{\alpha}}{2 \pi k_b T_{\alpha}(0)}\bigg]^{3/2} \exp{-\frac{r^2}{2 \sigma_{\alpha}(t)^2} - \frac{m_{\alpha}v^2}{2 k_b T_{\alpha}(t)}}.
\end{equation}
The radial distribution is a symmetric Gaussian distribution function with a deviation $\sigma_{\alpha}(t)$ is a function of time,
\begin{equation}
    \sigma^2_{\alpha}(t) = \sigma^2_{\alpha}(0)(1 + t^2/\tau_{exp}^2).
\end{equation}
The velocity distribution depends on the temperature $T_{\alpha}(t)$ of species $\alpha$,
\begin{equation}
    T_{\alpha}(t) = \frac{T_{\alpha}(0)}{1 + \frac{t^2}{\tau_{exp}^2}},
\end{equation}
Both $\sigma_{\alpha}(t)$ and $T_{\alpha}(t)$ depend on the characteristic expansion time $\tau_{exp}$
\begin{equation}
    \tau_{exp} = \sqrt{\frac{m_{\alpha} \sigma^2_{\alpha}(0)}{k_b[T_e(0) + T_i(0)]}}.
\end{equation}
Integrating velocity space, we can write the number density $\eta(t,\vb{r})$ as a function of time and radius from center of the sphere, 
\begin{equation}
    \eta(t,\vb{r}) = \eta_0 \bigg[\frac{T_{\alpha}(t)}{T_{\alpha}(0)}\bigg]^{3/2}e^{\frac{-r^2}{2 \sigma_{\alpha}^2}}.
\end{equation}
This derivation was used in creating Fig. \ref{fig:plasma_ball_dis}.

\subsection{Two Stream Instability}
\begin{figure}
    \centering
    \includegraphics[draft=false,scale=0.55]{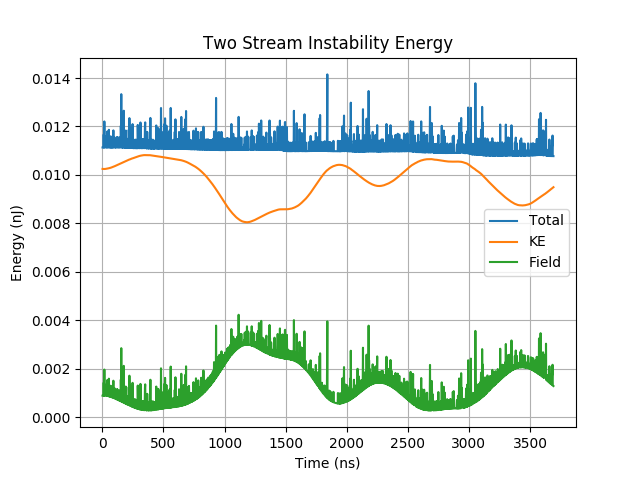}
    \caption{Energy plot for two stream instability with 3D periodic boundary conditions.}
    \label{fig:two_stream_energy}
\end{figure}

Next, we present the two stream instability test case from Birdsall and Langdon \cite{birdsall2018plasma}.
We shall follow its prescription and paraphrase it here.
Two streams of particles are setup, each with a density of $n_0$ and the same sign.
This test case was chosen due to analytic solutions associated with the growth rate of the instability.
While it is inherently a 1-D problem and has been done many times before by others, it is a classic problem that can adapted to 3-D and provides insight into code performance. 
Typically, this test is performed by using periodic boundary conditions.
Alternatively, when that functionality is not available, one can use perfectly electric conductor (PEC),
pack the volume with charge with the two velocity setup,
and emit particles from both PEC surfaces with the same velocity, such that you get a similar system. 
PEC provides image charge and thus retains the same physical features of the problem in a physically realizable configuration.  
The specific example parameter details can be found in Table \ref{table:two_stream}.
\begin{table} 
    \centering
    \caption{Two Stream Instability Parameters}
    \begin{tabular}{c|c}\label{table:two_stream}
         \textit{Parameter} & \textit{Value}  \\
         \hline
         Boundary Conditions & Periodic \\
         Solver Type & electrostatic \\
         species type & electrons \\ 
         density & $4.5 \cdot 10^{9} cm^{-3}$\\
         $v_0$ & $5 \cdot 10^{5}$ m/s \\
         Mesh Edge Length & 5 cm \\
         Number particles & 1000 \\
         macro-particle size & 9000.0 \\
         length of domain & 2 m\\
    \end{tabular}
\end{table}
 \begin{figure}
    \centering
    \includegraphics[draft=false,scale=0.55]{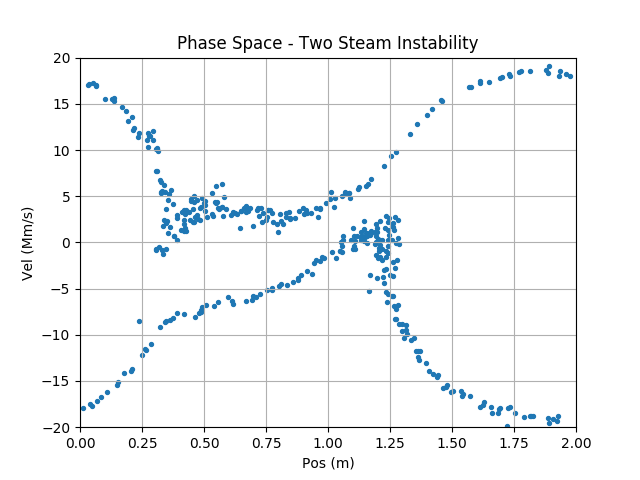}
    \caption{Two stream phase diagram for electrostatic finite element solver at 184 ns into the simulation with 400 particles.}
    \label{fig:phase_two_stream}
\end{figure}
The two-stream example shown here produces the standard swirls on the phases diagram shown in Fig. \ref{fig:phase_two_stream}. 
The energy of the system, shown in Fig. \ref{fig:two_stream_energy}, is conserved. 
The small bumps in energy in the field solve is caused from electrons moving too close to one another causing spikes in electric field values.

\section{Summary}
In this paper, we have provided four tests that can be used to validate 3D PIC codes. These tests have been developed as (a) there is significant recent interest in developing PIC codes to fully resolve complex
boundaries, as properties of these boundaries, and (b) as there is a void in examples with details that can be used to validate these codes. In all examples that we have developed, there exist quasi-analytic
solutions, as well as open source codes that can be used for benchmarking results. We have also provided sufficient detail so as to facilitate the user to examine all possible metrics that may subtly affect the
physics.

\ifCLASSOPTIONcompsoc
  \section*{Acknowledgments}
\else
  \section*{Acknowledgment}
\fi

This work was supported by SMART Scholarship program. We thanks the MSU Foundation for support through the Strategic Partnership Grant.
The authors would also like to thank the HPCC Facility,
Michigan State University, East Lansing, MI, USA.
\ifCLASSOPTIONcaptionsoff
  \newpage
\fi

\bibliography{fempic}{}
\bibliographystyle{ieeetr}







\end{document}